\documentclass{webofc}
\usepackage[varg]{txfonts}   
\usepackage{booktabs}
\usepackage{siunitx}
\usepackage{lineno}
\newcommand{\MET}{\ensuremath{p_\textrm{T}^\textrm{miss}}}
\newcommand{\mus}{\micro\second}
\usepackage{xcolor}
\begin{document}
\title{The CMS Trigger Upgrade for the HL-LHC}
%
%

\author{\firstname{Thiago Rafael} \lastname{Fernandez Perez Tomei}
\inst{1}
\fnsep\thanks{\email{Thiago.Tomei@cern.ch}}
on behalf of the CMS Collaboration
}

\institute{Universidade Estadual Paulista -- Unesp}

\abstract{%
  The CMS experiment has been designed with a two-level trigger system: the Level-1 Trigger, implemented on custom-designed electronics, and the High Level Trigger, a streamlined version of the CMS offline reconstruction software running on a computer farm.
  During its {second phase} the LHC will reach a luminosity of $7.5\times10^{34}\,\textrm{cm}^{-2}\,\textrm{s}^{-1}$ with a pileup of 200 collisions, {producing integrated luminosity greater than} 3000 fb$^{-1}$ over the full experimental run.
  To fully exploit the higher luminosity, the CMS experiment will introduce a more advanced Level-1 Trigger and increase the full readout rate from 100 kHz to 750 kHz.
  CMS is designing an efficient data-processing hardware trigger that will include tracking information and high-granularity calorimeter information.
  The current Level-1 conceptual design is expected to take full advantage of advances in FPGA and link technologies over the coming years, providing a high-performance, low-latency {system} for large throughput and sophisticated data correlation across diverse sources. 
  The higher luminosity, event complexity and input rate present an unprecedented challenge to the High Level Trigger that aims to achieve a similar efficiency and rejection factor as today despite the higher pileup and more pure preselection.
  In this presentation we will discuss the ongoing studies and prospects for the online reconstruction and selection algorithms for the high-luminosity era.
}
\maketitle
\section{The HL-LHC and the CMS Phase-2 Upgrade}

Full exploitation of the LHC remains the highest priority of the European Strategy for Particle Physics.
The High-Luminosity LHC (HL-LHC)~\cite{Apollinari:2015bam} is the natural upgrade path for the accelerator and was approved by the CERN Council in 2015.
The \emph{nominal (ultimate) configuration} of the accelerator will lead to pp collisions at the design energy of $\sqrt{s}$ = 14~TeV, and an instantaneous luminosity of {up to} 5.0 (7.5)~$\times$10\textsuperscript{34}~{(see Tab.~\ref{tab:trfpt-newTriDAS})}.
At nominal (ultimate) luminosity, the \emph{pileup} (occurrence of multiple pp interactions in the same or neighbouring bunch crossings) will reach an average of $\langle \textsf{PU} \rangle$ = 140 (200).
Those will be the harshest conditions to date in a hadron collider, with an experimental difficulty similar to that experienced in the Tevatron--LHC transition.
With that configuration, the accelerator will be able to deliver an integrated luminosity of 3000~fb\textsuperscript{-1} by 2035.

Coterminous with the HL-LHC era, the CMS Phase-2 Upgrade will bring the CMS detector capabilities up to the task~\cite{chatrchyan:2008aa,CMSCollaboration:2015zni}.
Its goal is to maintain the experimental performance in efficiency, resolution and background rejection for all physics observables.
One of the key CMS Phase-2 improvements will be the installation of an all-new tracker, which will have the capability of reading out \emph{stubs} -- matched pairs of hits on each side of a double layer, compatible with the passage of particles above a $p_\textrm{T}$ threshold -- at the bunch crossing frequency of 40~MHz.
{Those stubs will be input components of the L1 Track Trigger capability as described in Sec.~\ref{sec:trfpt-l1tu}.}
The barrel calorimeters (ECAL and HCAL) will have upgraded electronics, allowing 
full granularity readout at 40~MHz and compatibility with a L1 triggering rate of 750~kHz.
CMS will also have an all-new High Granularity Endcap Calorimeter (HGCAL), {substituting the current endcap sections of both ECAL and HCAL.}
{The HGCAL will be able to withstand the HL-LHC radiation dose and allow longitudinal layer-by-layer readout.}
A MIP Timing Detector (MTD) will be installed between the Tracker and Calorimeter systems, {providing precise timing measurement of charged particle tracks for an additional tool for pileup mitigation.}
For in-depth information of the CMS Phase-2 upgrade, we refer the reader to the CMS subsystems upgrade TDRs\footnote{All CMS reports are available at \texttt{http://cds.cern.ch/collection/CMS Reports}.}.

In order to address the challenging HL-LHC conditions, the CMS Trigger and Data Acquisition System -- TriDAS -- will also have to undergo an upgrade~\cite{CMSCollaboration:2283193}.
The Trigger part of the system will still be divided in a Level-1 Trigger (L1T), implemented in a set of FPGA boards, and a High-Level Trigger (HLT) implemented as a suite of algorithms running in an online farm of commercial processors.
Table~\ref{tab:trfpt-newTriDAS} gives some of the operating parameters of the new TriDAS, 
{based both on
linear extrapolations of CMS Phase-1 measurements of the CPU time requirements
and size of simulated Phase-2 events.}
Even assuming a rate reduction of 1/100 at the HLT, the online farm would need to have 18 times (27 times) more processing power (storage capacity) to maintain the current system performance.
A simple upscaling of the current paradigm is not cost-effective and innovative solutions must be pursued.

\begin{table}[htbp]
	\footnotesize
   \centering
\scalebox{0.925}{
\begin{tabular}{@{} lccc @{}} 
     \toprule
        & LHC  & \multicolumn{2}{c}{HL-LHC} \\
CMS detector    & Run 2 & \multicolumn{2}{c}{Phase-2} \\
\cmidrule{2-4}
Peak $\langle\textsf{PU}\rangle$ & 60 & 140 & 200 \\
\midrule
L1 accept rate (maximum)   & 100 kHz & 500 kHz & 750 kHz \\
Event Size & 2.0 MB &  5.7 MB &  7.4 MB \\
Event Network throughput   & 1.6 Tb/s &  23 Tb/s   & 44 Tb/s \\
Event Network buffer (60~s) &  12 TB &  171 TB & 333 TB \\
HLT accept rate &  1 kHz &  5 kHz &  7.5 kHz \\
HLT computing power &  0.5 MHS06 &  4.5 MHS06 &  9.2 MHS06 \\
Storage throughput & 2.5 GB/s &  31 GB/s &  61 GB/s \\
Storage capacity needed (1 day) &  0.2 PB &  2.7 PB &  5.3 PB \\
   \bottomrule
   \end{tabular}
   }
   \caption{During the HL-LHC era, the CMS Trigger and Data Acquisition system will be operating under much more difficult conditions than the original design,
   {both for the nominal ($\langle \textsf{PU} \rangle$ = 140) and ultimate ($\langle \textsf{PU} \rangle$ = 200) scenarios.}
   Both the system's processing power and its storage capacity will have to increase by an order of magnitude to face that new challenge.
   Table adapted from Ref.~\cite{CMSCollaboration:2283193}.}
   \label{tab:trfpt-newTriDAS}
\end{table}
\vspace{-0.3cm}

\section{The Level-1 Trigger Upgrade}
\label{sec:trfpt-l1tu}

\begin{figure}[htbp]
\centering
\includegraphics[width=0.67\textwidth,clip]{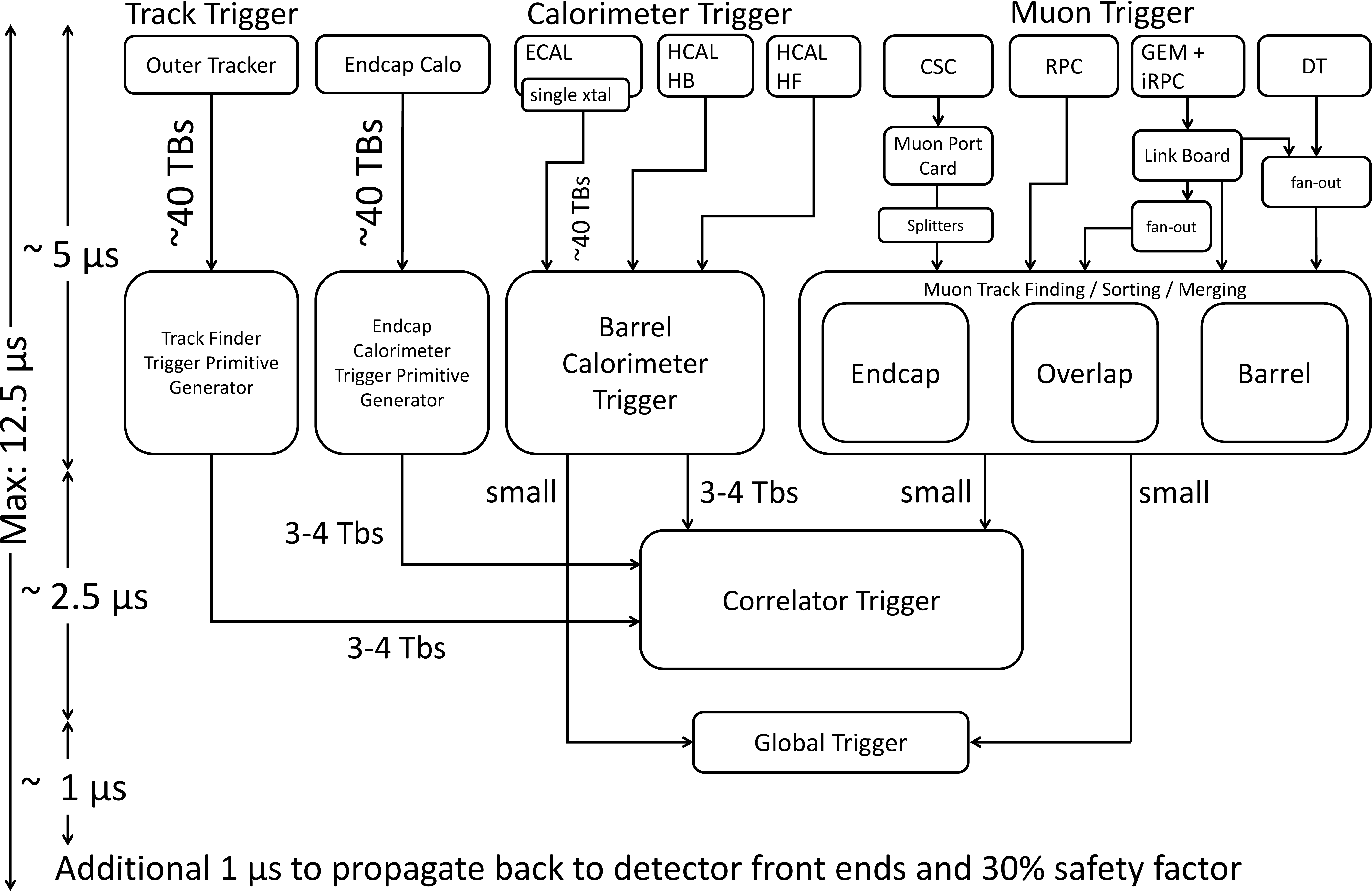}
\caption{The conceptual schematics for the CMS Phase-2 Level-1 Trigger System.
It will receive input from a number of CMS subdetectors:
Outer Tracker,
Barrel Calorimeters,
Endcap Calorimeter and the
Muon Systems.
Figure adapted from Ref.~\cite{CMSCollaboration:2283193}.
}
\label{fig:L1TConceptual}       
\end{figure}

A simplified scheme of the upgraded L1T can be seen in Fig.~\ref{fig:L1TConceptual}, and a detailed description is available in~\cite{Collaboration:2283192}.
The overall latency afforded for the system is \SI{12.5}{\mus}, and its maximum output rate is set at 750~kHz.
The Outer Tracker, the three calorimeter systems and the muon systems will all provide input to the L1T.

Tracker input at the L1T will be fundamental to achieve the target output rate, and the \emph{L1 Track Trigger} has two processing layers to deliver that: the
DAQ, Trigger and Control (DTC) layer
and the
Track Finding Processor (TFP) layer.
The DTC is responsible for the control and readout of the front-end modules.
It handles both the  main data stream to the central DAQ as well as the stub stream to the TFPs.
The track-finding algorithm uses the stubs as input and executes a hybrid algorithm that combines a
tracklet seed and road search algorithm
with a
Hough Transform and 3D Kalman filter tracking approach~\cite{Aggleton:2017ljc,Bartz:2019dkp}.

\begin{figure}[htbp]
\centering
\includegraphics[width=8.5cm,clip]{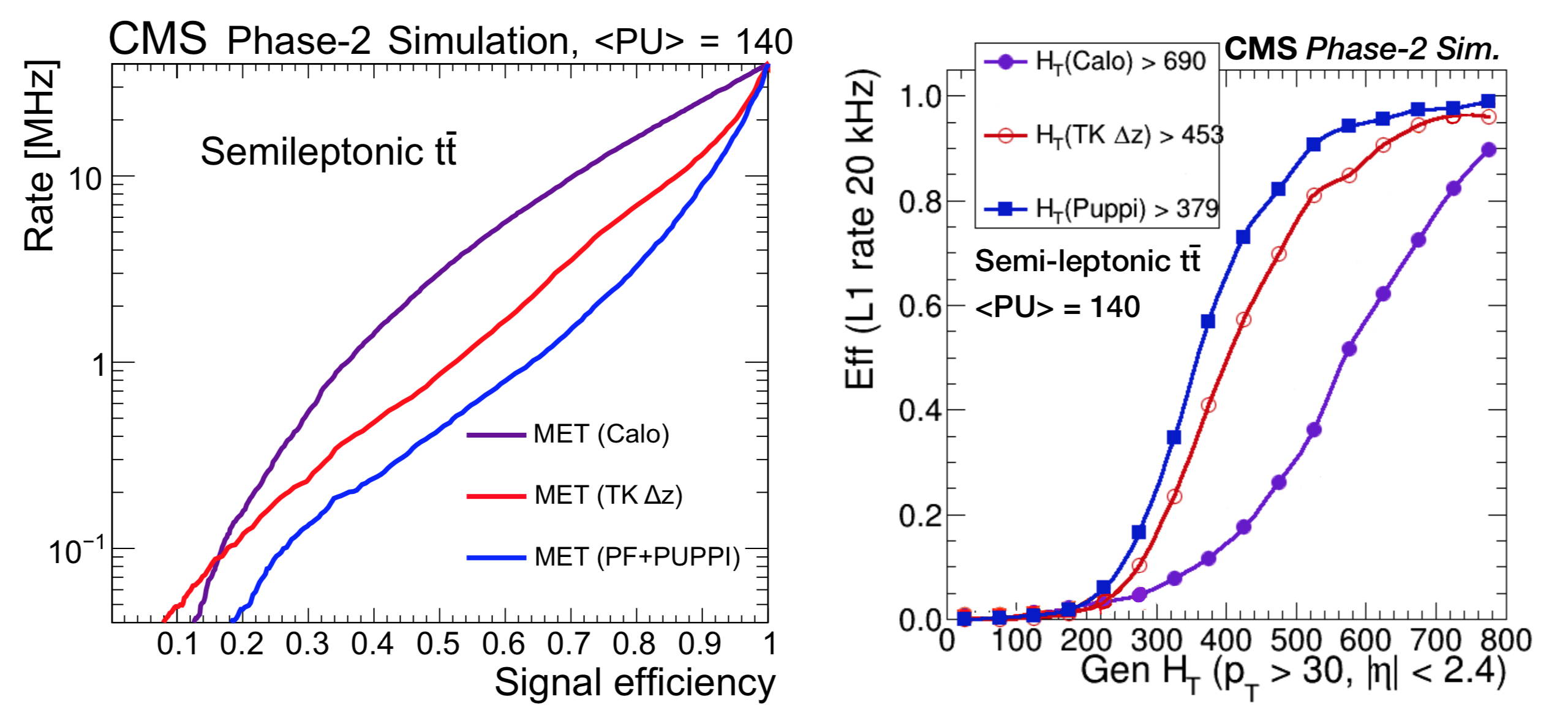}
\caption{Particle Flow and advanced pileup mitigations will allow for better Level-1 trigger algorithms.
This is illustrated by a study of the semileptonic $\textrm{t}\bar{\textrm{t}}$ process with different reconstruction methods:
{calorimeter-only;
track-only using tracks consistent with the primary vertex (TK $\Delta$z);
and PF+PUPPI.}
On the left is the rate as a function of efficiency curve (MET stands for missing transverse momentum, same as \MET{}.)
On the right is the turn-on curve for a fixed Level-1 rate of 20~kHz.
Figure from Ref.~\cite{Collaboration:2283192}.
}
\label{fig:PUPPIatL1}       
\end{figure}

The presence of the track trigger and the improved calorimetry system, together with the advent of more powerful FPGAs, will allow to perform \emph{Particle Flow} (PF) reconstruction at Level-1 during CMS Phase-2.
That in turn will open up the possibility of identifying the majority of the particles produced by the collisions and deploying advanced pileup mitigation techniques like PileUp Per Particle Identification (PUPPI), which assigns weights to the reconstructed particles roughly proportional to their probability of coming from a PU interaction.
Figure~\ref{fig:PUPPIatL1} shows the improvement using PF+PUPPI on event energy sum quantities (\MET{} and $H_\textrm{T} \equiv$ scalar $p_\textrm{T}$ sum of all jets in the event) when compared to calorimeter-only reconstruction and to a simpler usage of track trigger information.

\clearpage

The Level-1 Trigger Technical Design Report, an update on the results presented on the Interim Document,
{was released on June 2020~\cite{collaboration:2714892}.}
For completeness, we reproduce in Table~\ref{tab:prelimL1Menu} the L1 menu that was publicly available at the time of this conference.
The final report showcases the enhanced physics performance of the upgraded L1T.
It includes the presence of leptonic paths with extended pseudorapidity coverage, paths for reconstruction of light mesons and of displaced objects, paths that make usage of machine learning and of timing information, amongst other improvements.

\begin{table}[htbp]
\centering
\caption{Preliminary Level-1 Trigger menu, adapted from Ref.~\cite{Collaboration:2283192}.
{Objects which demand L1 tracking are indicated with ``(tk)''.}
{Rates for e/$\gamma$ (EG) algorithms are calculated in barrel only, indicated with an asterisk.}
{The last line of the table represents an extrapolation of this partial menu to the total rate required for a complete physics menu, including endcap EG and \MET{} algorithms.}
}
\label{tab:prelimL1Menu}       
\scalebox{0.925}{
\footnotesize
\begin{tabular}{lccc}
\toprule
$L=5.6 \times 10^{34} \mathrm{cm}^{-2} \mathrm{s}^{-1}, \quad\langle \textsf{PU}\rangle= 140$ & \multicolumn{3}{c}{L1 trigger}\\
$L=8.0 \times 10^{34} \mathrm{cm}^{-2} \mathrm{s}^{-1}, \quad\langle \textsf{PU}\rangle= 200$ & \multicolumn{3}{c}{with L1 tracks}\\
\midrule
 & & & Offline\\
Trigger   & \multicolumn{2}{c}{Rate} & threshold(s)\\
algorithm & \multicolumn{2}{c}{[kHz]} & [GeV]\\
\midrule
$\langle PU \rangle$ & 140 & 200 &\\
\midrule
Single Mu (tk)                     	& 14 & 27 & 18\\
Double Mu (tk)                     	& 1.1 & 1.2 & 14, 10\\
Ele$^*$ (iso tk) + Mu (tk)         	& 0.7 & 0.2 & 19, 10.5\\
Single Ele$^*$ (tk)					& 16 & 38 & 31\\
Single iso Ele$^*$ (tk) 				& 13 & 26 & 27\\
Single $\gamma^*$ (tk-iso) 			& 31 & 19 & 31\\
Ele$^*$ (iso tk) + e / $\gamma^*$  	& 11 & 7.3 & 22, 16\\
Double $\gamma^*$ (tk-iso) 			& 17 & 5 & 22, 16\\
Single Tau (tk) 					& 13 & 38 & 88\\
Tau (tk) + Tau						& 32.& 55 & 56, 56\\
Ele$^*$ (iso tk) + Tau				& 7.4 & 23 & 19, 50\\
Tau (tk) + Mu (tk)					& 5.4 & 6 & 45, 14\\
Single Jet							& 42 & 69 & 173\\
Double Jet (tk)						& 26 & 43 & 2@136\\
Quad Jet (tk)						& 12 & 45 & 4@72\\
Single Ele$^*$ (tk) + Jet			& 15 & 15 & 23, 66\\
Single Mu (tk) + Jet 				& 8.8 & 12 & 16, 66\\
Single Ele$^*$ (tk) + $H_\textrm{T}^\textrm{miss}$ (tk) & 10 & 45 & 23, 95\\
Single Mu (tk) + $H_\textrm{T}^\textrm{miss}$ (tk) & 2.7 & 8 & 16, 95\\
$H_\textrm{T}$ (tk) & 13 & 24 & 350 \\
\midrule
Rate for above triggers$^*$ & 180 & 305 &\\
Est. rate (full EG $\eta$ range) & & 390 &\\
\midrule
\textbf{Est. total L1 menu rate ($\times$ 1.3)} & \textbf{260} & \textbf{500}&\\
\bottomrule
\end{tabular}
}
\end{table}
\vspace{-0.3cm}

\section{The High-Level Trigger Upgrade}
\label{sec:trfpt-hltu}

The HLT upgrade is slated to deliver a working system by the beginning of Run~4 in 2027.
That system will have to deal not only with the increased pileup conditions but with an input rate that is {up to} 7.5 times larger.
Additionally, the CMS Phase-2 subdetectors are much more complicated than their current counterparts; the HGCAL comprises close to six million channels, to be compared with approximately 236,000 channels of the full calorimetry system at the start of the experiment.
{That complexity leads to a DAQ challenge; the detector readout will need an interconnect capable of 44~Tb/s throughput, as mentioned in Tab.~\ref{tab:trfpt-newTriDAS}.}

\begin{figure}[ht]
\centering
\sidecaption
\includegraphics[width=7.2cm,clip]{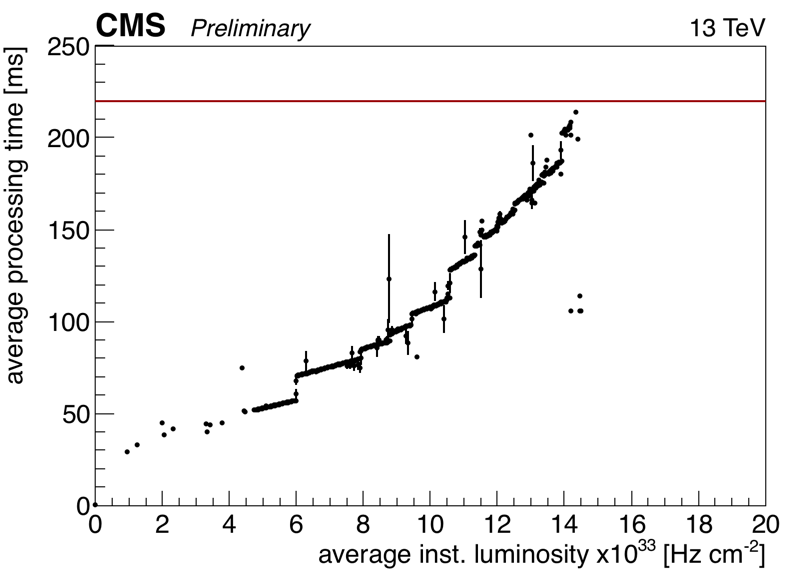}
\caption{The average processing time of the HLT grows faster than linearly with the instantaneous luminosity.
The red line marks the capacity of the HLT farm in 2016.
{The breakdown of the CPU time spent in the reconstruction code during Run 2 conditions is approximately:
10\% in ECAL, 10\% in HCAL, 35\% in tracking and 20\% in Particle Flow~\cite{CMSCollaboration:2283193}.}
Figure available from~\cite{plotsTiming}.
}
\label{fig:HLTtiming}       
\end{figure}

The timing of the algorithm suite remains one of the most difficult issues to be addressed.
Figure~\ref{fig:HLTtiming} shows that the dependency of the average processing time of the HLT with the average instantaneous luminosity grows faster than linearly.
Studies are ongoing to understand the dependency with average pileup and with pileup density (vertices/cm).

\begin{figure}[ht]
\centering
\sidecaption
\includegraphics[width=7.2cm,clip]{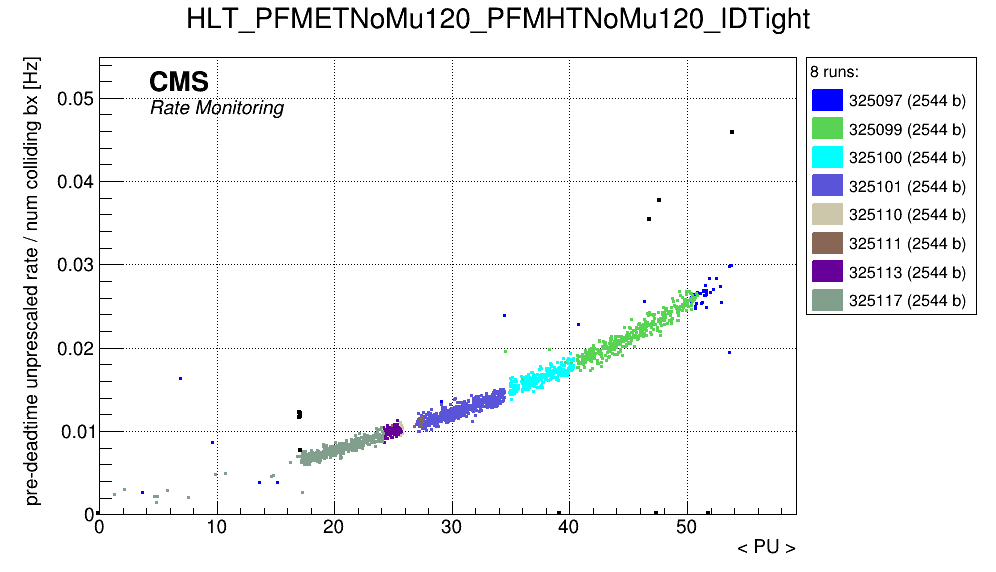}
\caption{The data-taking rate of an algorithm that selects events with $\MET{} >$ 120~GeV grows faster than linearly with the average event PU. The different colored sets of points represent different experimental runs in a single LHC fill with 2544 bunches.}
\label{fig:METrate}       
\end{figure}

The evolution of the absolute rate of the HLT system to Phase-2 conditions is also being studied.
The architecture of the system is such that the rate is distributed amongst many different data-taking paths.
Algorithms responsible for acquiring events where a single lepton (electron or muon) was produced represent approximately 25\% of the HLT rate;
those scale almost linearly with the pileup. On the other hand, the rate of \MET{} algorithms again grows faster than linearly with the pileup, as can be seen in Fig.~\ref{fig:METrate}.

From a computing point of view, there are a few approaches to tackle those challenges.
The fact that CMSSW is fully multithreaded allows for full usage of the multicore systems that have become the norm in computing.
Additionally, code modernisation and streamlining campaigns help uncover inefficient sections of the algorithm suite.
{Finally, it is likely that the hardware available in the market at the time of the HLT cluster purchase will have a better price/performance ratio compared to the offers available now.}
From 2008 to 2018, this improvement assumption was approximately true for computing nodes equipped with Intel\textsuperscript{\textregistered{}} processors:
\SI{14}{\nano\meter}, 2018 processor-equipped nodes are ten times more powerful, in terms of the HEP-SPEC06 benchmark, than \SI{45}{\nano\meter}, 2008 processor nodes\footnote{The 2018 processors in question have 16 processing cores, whilst the 2008 processors had only four.}.

{An additional strategy that allows a higher data-taking rate is the the usage of the \emph{scouting} technique.}
This approach gives up on recording the full raw data from the detector; instead, one saves only the output objects of the online reconstruction at the desired level.
The limitation of scouting with calorimetric-based objects is essentially given by the L1 reconstruction thresholds.
On the other hand, scouting with full particle-flow objects is limited by the HLT timing budget.
Both approaches have already been successfully used at CMS~\cite{Khachatryan:2016ecr,Sirunyan:2019pnb,Sirunyan:2019wqq}; a new approach for scouting with L1-based objects was recently proposed~\cite{L1.talk.at.CHEP}.

The deployment of \emph{heterogeneous architectures} at the High-Level Trigger is yet another approach to tackle the HL-LHC data-taking challenge.
The combination of specialised hardware (GPUs, FPGAs, ARM cores, etc.) with standard x86-64 CPUs allows to offload some tasks to those accelerators.
{This problem is currently addressed at CMS, with a successful demonstration of offloading the pixel reconstruction to GPUs, and a preliminary implementation scheduled to be deployed in Run~3, by the Patatrack project~\cite{CMS-DP-2018-059}.}


\section{Conclusions}

The High-Luminosity LHC will bring the harshest conditions in a high-energy collider experiment to date.
The jump in computing needs will be even bigger than at the start of the LHC era.
CMS is undergoing a full upgrade program in order to meet the challenge, with completely new subsystems (HGCAL, MTD) being built and enhancements being made to the existing ones.
Trigger and Data Acquisition remains one of the hardest problems to tackle, due to the sheer amount and the complexity of the events to be collected.
Sophisticated reconstruction at Level-1, with Track Trigger, Particle Flow and pileup mitigation techniques is being developed.
At the High-Level Trigger, the deployment of heterogeneous hardware (FPGAs, GPUs), the optimisation of the reconstruction algorithms, and the adoption of alternative data-taking strategies like scouting are the keys to success.
Overcoming this challenge is mandatory to unlock the physics potential of the 3000~fb\textsuperscript{-1} of collision data that will be delivered by the HL-LHC.

\end{document}